# A Simple Walk Model for Reproducing Power Laws in Human mobility

Shuji Shinohara[a,*], Daiki Morita[a], Hayato Hirai[a], Ryosuke Kuribayashi[a], Nobuhito Manome[b,c], Toru Moriyama[d], Yoshihiro Nakajima[e], Yukio-Pegio Gunji[f], and Ung-il Chung[b]

[a] *School of Science and Engineering, Tokyo Denki University, Saitama, Japan*

[b] *Department of Bioengineering, Graduate School of Engineering, The University of Tokyo, Tokyo, Japan*

[c] *Department of Research and Development, SoftBank Robotics Group Corp., Tokyo, Japan*

[d] *Faculty of Textile Science, Shinshu University, Ueda, Japan*

[e] *Graduate School of Economics, Osaka City University, Osaka, Japan*

[f] *Department of Intermedia Art and Science, School of Fundamental Science and Technology, Waseda University, Tokyo, Japan*

\* Corresponding author

E-mail: s.shinohara@mail.dendai.ac.jp

Postal address: School of Science and Engineering, Tokyo Denki University, Ishizaka, Hatoyama-machi, Hiki-gun, Saitama 350-0394, Japan

Acknowledgments

This study was supported by JSPS KAKENHI [Grant No. JP21K12009].

.

# Abstract

Identifying statistical patterns characterizing human trajectories is crucial for public health, traffic engineering, city planning, and epidemic modeling. Recent developments in global positioning systems and mobile phone networks have enabled the collection of substantial information on human movement. Analyses of these data have revealed various power laws in the temporal and spatial statistical patterns of human mobility. For example, jump size and waiting time distributions follow power laws. Zipf's law was also established for the frequency of visits to each location and rank. Relationship $S(t) \sim t^\mu$ exists between time $t$ and the number of sites visited up to $S(t)$. Recently, a universal law of visitation for human mobility was established. Specifically, the number of people per unit area $\rho(r,f)$, who reside at distance $r$ from a particular location and visit that location $f$ times in a given period, is inversely proportional to the square of rf, i.e., $\rho(r,f) \propto (rf)^{-2}$ holds. The exploration and preferential return (EPR) model and its improved versions have been proposed to reproduce the above scaling laws. However, some rules that follow the power law are preinstalled in the EPR model. We propose a simple walking model to generate movements toward and away from a target via a single mechanism by relaxing the concept of approaching a target. Our model can reproduce the abovementioned power laws and some of the rules used in the EPR model are generated. These results provide a new perspective on why or how the scaling laws observed in human mobility behavior arise.

# 1. Introduction

Lévy walks have been observed in the migratory behavior of organisms across a range of scales, from bacteria and T cells to humans [1–5]. These walks, a specialized type of random walk, exhibit step lengths $l$ that follow power law distribution $P(l) = al^{-\mu}, 1 < \mu \leq 3$, in contrast to the exponentially distributed step lengths of the Brownian walk (where the frequency of step length $l$ is characterized by an exponential distribution, $P(l) = \lambda e^{-\lambda l}$). Lévy walks are particularly notable for their occasional, long, and linear movements. Lévy walks with exponents close to two have been frequently documented in various organisms, sparking interest in the reasons for these patterns [1, 6–11]. Such walks, when the exponent is two, are also known as Cauchy walks. The Lévy flight foraging hypothesis (LFFH) [12, 13] suggests that under conditions where food is scarce and randomly dispersed and predators lack any memory of food locations, Cauchy walks represent the most efficient foraging strategy and offer evolutionary benefits [14]. As highlighted in the LFFH, Lévy walks are not universally applicable across all environments or conditions. Humphries et al. [4] found that Lévy behavior is associated with environments where prey is sparse, whereas Brownian movements correlate with areas that have abundant prey. Additionally, Huda et al. [12] demonstrated that metastatic cells exhibit Lévy walks, whereas non-metastatic cancer cells engage in simple diffusive movements. Shinohara et al. [15] proposed a walk model that continuously generates a Brownian walk to a Cauchy walk in a multidimensional space by varying the control parameters.

In general, human mobility behavior is characterized by a high frequency of returning to and staying in specific familiar places, such as the home and workplace, rather than random walking as in the foraging behavior of animals. Uncovering statistical patterns unique to human mobility has crucial implications for public health, traffic engineering, city planning, and epidemic modeling [16–19].

Human mobile behavior has been previously analyzed using tracking data from bank notes [20]. In recent years, the development of global positioning systems (GPS) and cell phone networks has enabled the collection of substantial amounts of information on human movement throughout society, such as vehicle GPS trajectories and calling description records (CDR) [18, 21, 22]. The widespread use of smartphones has also enabled analysis using mobile flow records (MFR) [22]. MFR provides much higher time-resolved user locations and captures more detailed motion behavior than CDR. Analyses of these data revealed various scaling properties in the spatial and temporal statistical patterns of human trajectories.

For example, the distribution of the jump size $\Delta r$ exhibits power law $P(\Delta r) \sim \Delta r^{-\alpha}$ as in the case of various living organisms, where $\Delta r$ denotes the distance covered by an individual between consecutive sightings [14, 17, 22]. Similarly, the distribution of waiting time $\Delta t$ shows power law $P(\Delta t) \sim \Delta t^{-\beta}$, where $\Delta t$ denotes the time spent by an individual at a location [16, 17, 22]. As for the frequency of visits, the frequency $f_k$ of the $k_{th}$ most visited site follows Zipf's law $f_k \propto k^{-\xi}$ [16, 17, 23]. There exists relationship $S(t) \propto t^{\mu}$ between elapsed time $t$ and number of sites $S(t)$ visited up to that time $t$ [16, 17]. Some of the above power laws are observed not only in real space but also in virtual space, such as navigation in TV programs and online shopping sites [17]. Schläpfer et al. found a universal law of visitation for human mobility that links travel distance to travel frequency. Specifically, they found that the number of people per unit area, $\rho(r, f)$, who live at distance $r$ from a particular location and visit that location $f$ times in a given period is inversely proportional to the square of $rf$, i.e., $\rho(r, f) \propto (rf)^{-2}$ holds [24]. Surprisingly, this law holds universally in various cities worldwide, as analyzed in this study.

An exploration and preferential return (EPR) model is proposed to reproduce the various scaling laws described above [16]. In the EPR model, the agent performs one of two actions each time: exploring a new location or returning to a previously visited location. The agent visits a new location with probability $P_{new} = \rho S^{-\gamma}$, and revisits a previously visited location with probability $P_{ret} = 1 - P_{new}$, where $S$ indicates the number of sites visited. In the case of exploration, the direction of movement is random and jump size $\Delta r$ is sampled randomly from power distribution $P(\Delta r) \sim \Delta r^{-\alpha}$. In the case of return, the visited sites are selected probabilistically in proportion to the frequency of previous visits. The waiting time, $\Delta t$, is not constant but is sampled randomly from power distribution $P(\Delta t) \sim \Delta t^{-\beta}$. Recently, various improved models using this model as a platform have been proposed to reproduce empirical data more accurately [17, 24]. The EPR model can effectively reproduce the various statistical patterns found in the empirical data. However, as shown above, some rules that follow the power law are preinstalled in the model.

In this study, we simplified the walking model proposed by Shinohara et al. [15] and simulated human mobility using the simplified model. We showed that the model can reproduce some of the above power laws although it does not explicitly incorporate rules that follow the power law.

## 2. Methods

### 2.1 Walk Model

Let us assume that, at each time, the agent explores the area around the base point, $B = \begin{pmatrix} b_1 \\ b_2 \end{pmatrix}$. The base point can be the current position of the agent or a fixed point, such as home. The point that the agent wants to explore is called the target point and is denoted by $T = \begin{pmatrix} t_1 \\ t_2 \end{pmatrix}$. The direction $\theta$ of $T$ was sampled from the uniform distribution of $[0, 2\pi)$, and distance $d$ from $B$ to $T$ was sampled randomly from one-dimensional exponential distribution $P(d) = \lambda e^{-\lambda d}$. Finally, let $T = \begin{pmatrix} b_1 + d\cos\theta \\ b_2 + d\sin\theta \end{pmatrix}$. Let $X = \begin{pmatrix} x_1 \\ x_2 \end{pmatrix}$ denote the current position of the agent. The difference vector between $T$ and $X$ is denoted as $R = T - X = \begin{pmatrix} t_1 - x_1 \\ t_2 - x_2 \end{pmatrix} = \begin{pmatrix} r_1 \\ r_2 \end{pmatrix}$. The norm of the vector is denoted by $r = \|R\|$. The position of the agent at the next time point is $X_{next}$, and the movement vector is $X_{next} - X = \Delta X = \begin{pmatrix} \Delta x_1 \\ \Delta x_2 \end{pmatrix}$ (Fig. 1).

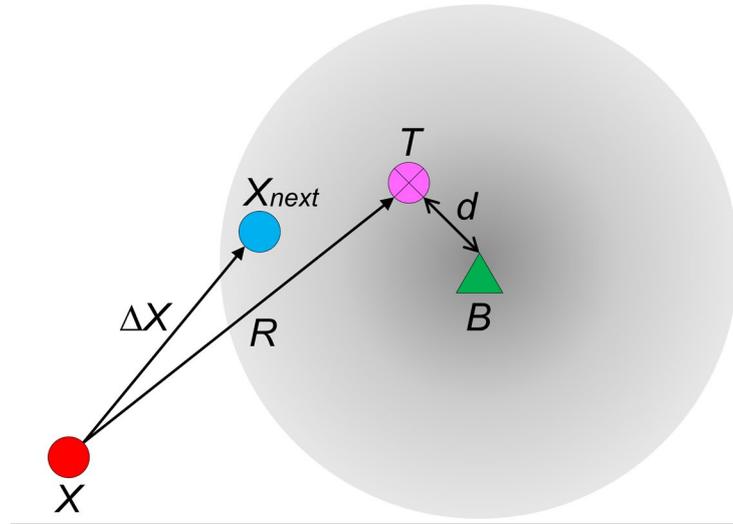

**Fig. 1** Diagram of the walk model. The agent decides on target point *T* in vicinity of base point *B* to explore around *B* and tries to approach it. However, the agent does not always reach *T*

Under the above settings, the agent attempts to approach target point *T* according to Equations (1) and (2) [15]:

$$\Delta x_i = \alpha \eta_i \frac{r^2}{r_i} \quad (1)$$

$$\eta_i = \beta_i^{1-|\gamma|} \left(\frac{r_i^2}{r^2}\right)^{\gamma} \quad (2)$$

Here, $0 \leq \alpha \leq 1$ denotes the rate of change. $\beta_i$ was set randomly, where $\beta_i$ satisfied the following conditions: $0 \leq \beta_i \leq 1$ and $\sum_{i=1}^{2} \beta_i = 1$.

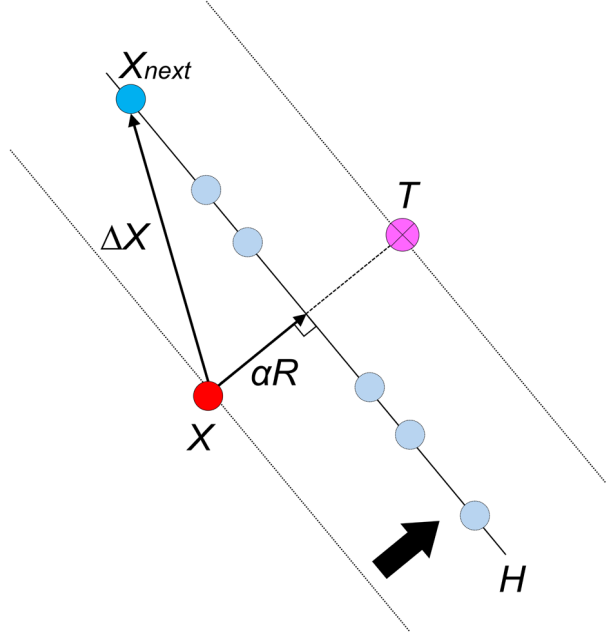

**Fig. 2** Rules for approaching the target point. When $\gamma = 0$, the agent moves to a point on hyperplane *H*, at next time. When $\gamma = 1$, the agent moves to one specific point on hyperplane *H*, i.e., intersection of *H* and *αR*

If $\gamma = 1$, then $\Delta X = \alpha R$ and the agent approaches *T* using the shortest path. In addition, if $\alpha = 1$, then $\Delta X = R$, i.e., $X_{next} = T$, and the agent moves directly from *X* to *T*. Furthermore, if the current position of the agent is the base point, i.e., *X=B*, the agent moves from *B* to *T* each time and its walking pattern is a Brownian walk. If $X = B$ and $\gamma$ vary from 1 to 0, the walking pattern changes continuously from a Brownian walk to a Cauchy walk [15]. In the case of $\gamma = 0$, from Equations (1) and (2), since $r_i \Delta x_i = \alpha \beta_i r^2$ and $\sum_{i=1}^{2} \beta_i = 1$ are satisfied, $R \cdot \Delta X = \sum_{i=1}^{2} r_i \Delta x_i = \sum_{i=1}^{2} \alpha \beta_i r^2 = \alpha r^2$ holds. In particular, $\Delta X$ refers to a point on hyperplane $H(R, \alpha r^2) = \{\Delta X \in \mathbb{R}^2 \mid R \cdot \Delta X = \alpha r^2\}$ whose normal vector is *R* (Fig. 2).

In this study, we simplify this model by fixing the parameters to $\alpha = 1$ and $\gamma = 0$, and redefining Equations (1) and (2) as a movement rule without parameters as follows:

$$\Delta x_i = \beta_i \frac{r^2}{r_i} \qquad (3)$$

Each time, the agent sets a random target point around the base point and updates the current position according to the movement rule expressed in Equation (3). However, the following points should be noted. The agent has the freedom to set the direction of each axis, resulting in each axis being randomly set each time. Let us denote the 2-dimensional standard basis as $\{e_1, e_2\}$. Here $e_i$ represents a 2-dimensional fundamental vector where the $i$-th element is 1 and all other elements are 0.

Initially, a new 2-dimensional orthonormal system, $\{e'_1, e'_2\}$, is generated using the Gram-Schmidt orthogonalization method. In this context, a relationship is established between the difference vector $R = \begin{pmatrix} r_1 \\ r_2 \end{pmatrix}$ in the standard basis and the difference vector $R' = \begin{pmatrix} r'_1 \\ r'_2 \end{pmatrix}$ in the new orthonormal system.

$$(e'_1 \; e'_2) \begin{pmatrix} r'_1 \\ r'_2 \end{pmatrix} = (e_1 \; e_2) \begin{pmatrix} r_1 \\ r_2 \end{pmatrix} \qquad (4)$$

If the transformation matrix between the two orthogonal systems is $A = (a_{ij})$ and $(e'_1 \; e'_2) = (e_1 \; e_2) A$, then $A = (e_1 \; e_2)^{-1} (e'_1 \; e'_2)$.

For the standard basis, $(e_1 \; e_2)$ and $(e_1 \; e_2)^{-1}$ are the identity matrices, and hence, $A = (e'_1 \; e'_2)$. From Equation (4), it can be observed that the relationship between $AR' = R$ and $R' = A^{-1}R$ can be established.

The specific procedure for calculating the movement vector is described below. Initially, a new random target point $T$ around the base point and a new 2-dimensional orthonormal system are generated. Next, using the transformation matrix, the difference vector $R$ in the standard basis is converted to the difference vector $R' = A^{-1}R$ in the new orthonormal system. Subsequently, Equation (3) is employed to derive the movement vector $\Delta X' = \begin{pmatrix} \Delta x'_1 \\ \Delta x'_2 \end{pmatrix}$ in the new orthonormal system. Next, $\Delta X = A \Delta X'$ is utilized to return $\Delta X'$ to the movement vector $\Delta X$ in the standard basis. Finally, the agent's current position is updated to $X_{next} = X + \Delta X$. This process is repeated iteratively.

## 2.2 Simulation Setting

In this study, the aforementioned walk model was used to simulate human mobility. The model generates a Cauchy walk when the base point is set to the current position of the agent. However, for this simulation, the base point is assumed to be home and fixed to $B = \begin{pmatrix} 0 \\ 0 \end{pmatrix}$. We assumed that the target point was very close to the base point and set $\lambda = 10^{20}$.

The area in which the agent could move was limited to the inside of a circle of radius 1.0 centered at the base point, i.e., the origin. The initial position of the agent was set at a random point within this region.

Assuming that the agent can move only at a finite speed, the maximum distance it can move per unit time was set to $\|\Delta X\| = 0.7$. The simulation period was set to 100000 steps. However, when the agent went outside the area of the circle of radius 1, the simulation was terminated. The simulation was run for 1000000 trials using different random seeds. In the analysis, the position of the agent was discretized by dividing the region into square sites of 0.01 per side.

## 3. Results

Figure 3(a) and (b) show examples of the movement trajectory of an agent and the time evolution of the distance from the base point to the agent.

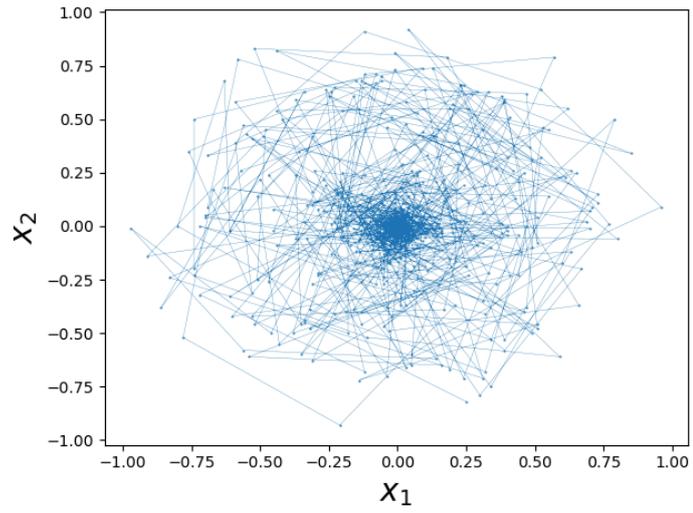

(a)

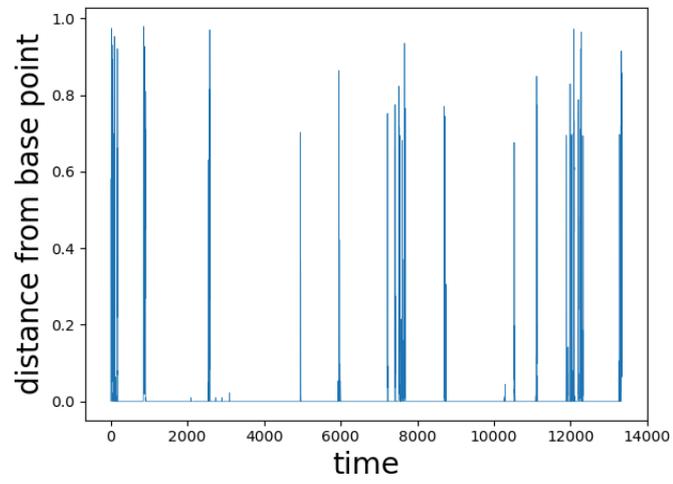

(b)

**Fig. 3.** Example of an agent's mobility. (a) Movement trajectory. (b) Time evolution of the distance from the base point to agent.

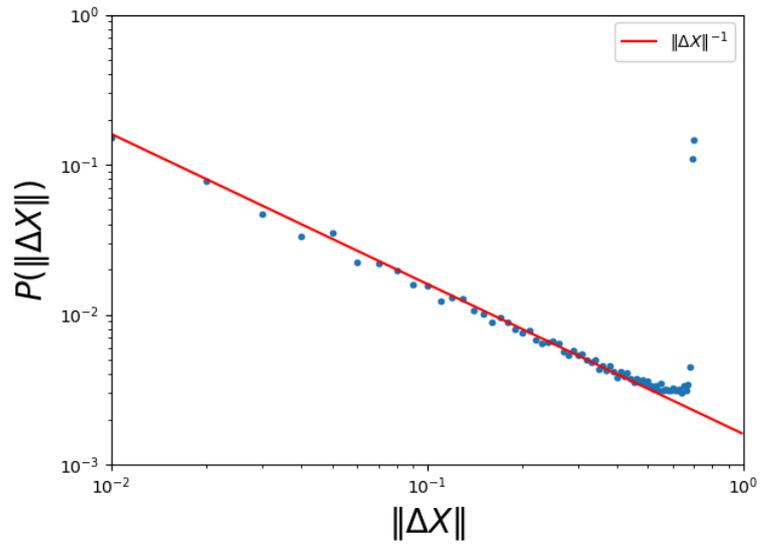

**Fig. 4.** Distance distribution $P(\|\Delta X\|)$. Both axes are displayed on the logarithmic scale. The red line represents $\|\Delta X\|^{-1.0}$

Figure 4 shows the distance distribution $P(\|\Delta X\|)$. The figure shows power law $P(\|\Delta X\|) \sim \|\Delta X\|^{-1.0}$ over a wide range. Note that the maximum jump size is limited to $\|\Delta X\| = 0.7$; thus, no jump size larger than 0.7 can appear.

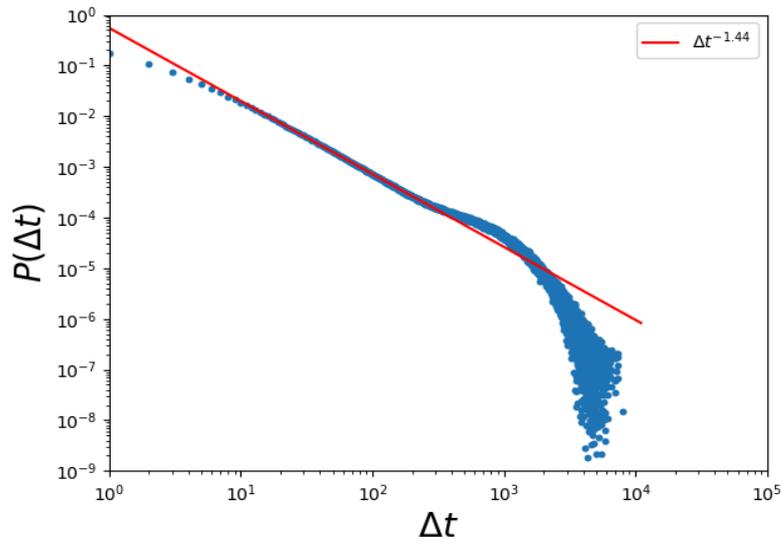

**Fig. 5** Waiting time distribution $P(\Delta t)$. Both axes are displayed on the logarithmic scale. The red line represents $\Delta t^{-1.44}$

Figure 5 shows the waiting time distribution $P(\Delta t)$. In this simulation, the waiting time is defined as the number of time steps spent consecutively at the same site. As shown in the figure, power law $P(\Delta t) \sim \Delta t^{-1.44}$ was observed for relatively short stays up to several hundred steps.

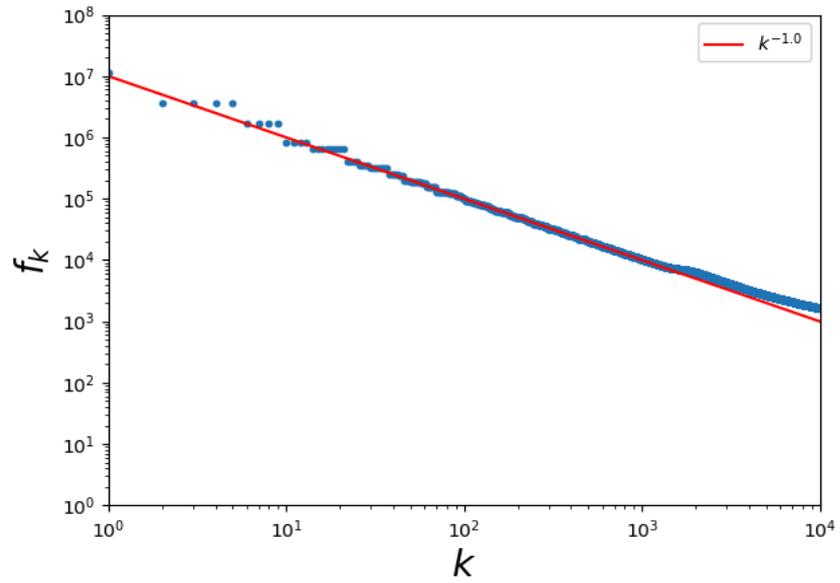

**Fig. 6** Zipf plot of the frequency of visits to each site. The horizontal axis represents a rank $k$ of sites in the order of number of visits, and vertical axis represents the frequency $f_k$ of the visits to site. Both axes are displayed on the logarithmic scale. The red line represents $k^{-1.0}$

Figure 6 illustrates a Zipf plot of the frequency of visits to each site. The horizontal axis represents the rank $k$ of the sites in the order of the number of visits, and the vertical axis represents the frequency $f_k$ of visits to the site. The figure shows that Zipf's law $f_k \propto k^{-1.0}$ holds over a wide range.

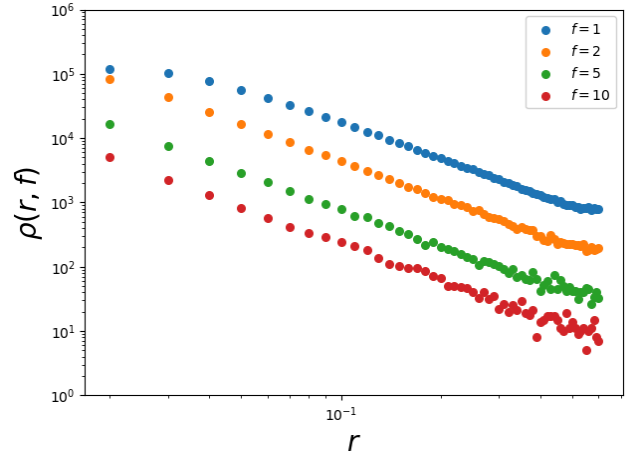

(a)

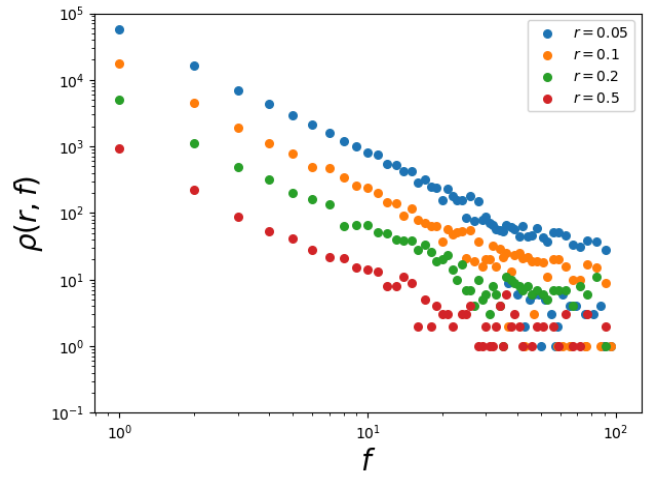

(b)

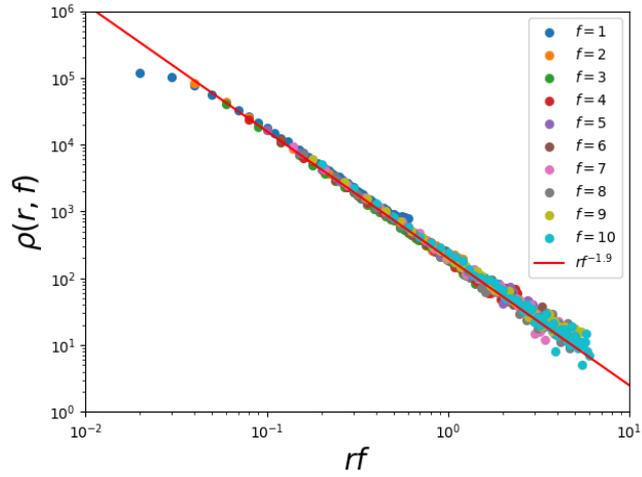

(c)

**Fig. 7** Number of agents $\rho(r,f)$ per unit area who reside at distance $r$ away from a particular location and visit it $f$ times during 1000 time steps. In (a), (b), and (c), the horizontal axes represent $f$, $r$, and $rf$, respectively, and the vertical axis represents $\rho(r,f)$. Both axes are displayed on the logarithmic scale. The red line represents $\rho(r,f) \propto (rf)^{-1.9}$ in Fig. 7 (c)

Figure 7(a), (b), and (c) show number of agents $\rho(r,f)$ per unit area who reside at distance $r$ away from a particular location and visit that location $f$ times in a given period. Here, the period is set to 1000 steps. As mentioned previously, the simulation ended when the agent moves outside the area of the circle of radius 1. Therefore, the simulation period was different in each trial. Let $\tau$ be the simulation period in a trial and assume that a location is visited $v$ times during that period. In this case, the number of visits to the location per 1000 steps was calculated as $f = \dfrac{1000v}{\tau}$. From the figures, it can be observed that power law $\rho(r,f) \propto (rf)^{-1.9}$ is established among $r$, $f$, and $\rho$.

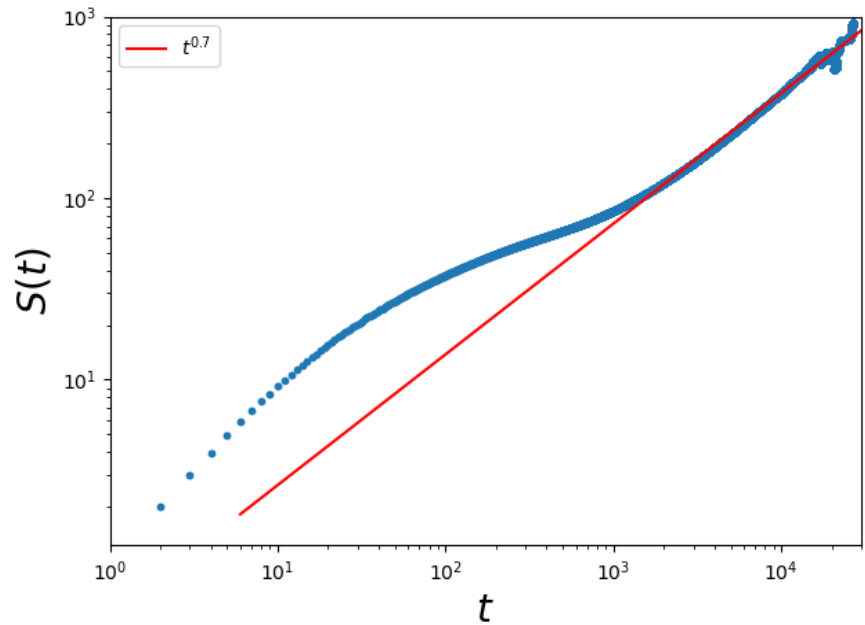

**Fig. 8** Relationship between time $t$ and number of visited sites $S(t)$. Both axes are displayed on the logarithmic scale. The red line represents $t^{0.7}$

Figure 8 illustrates the relationship between time $t$ and number of sites visited $S(t)$. The figure shows that power law $S(t) \propto t^{0.7}$ holds in the range of several thousand steps or more.

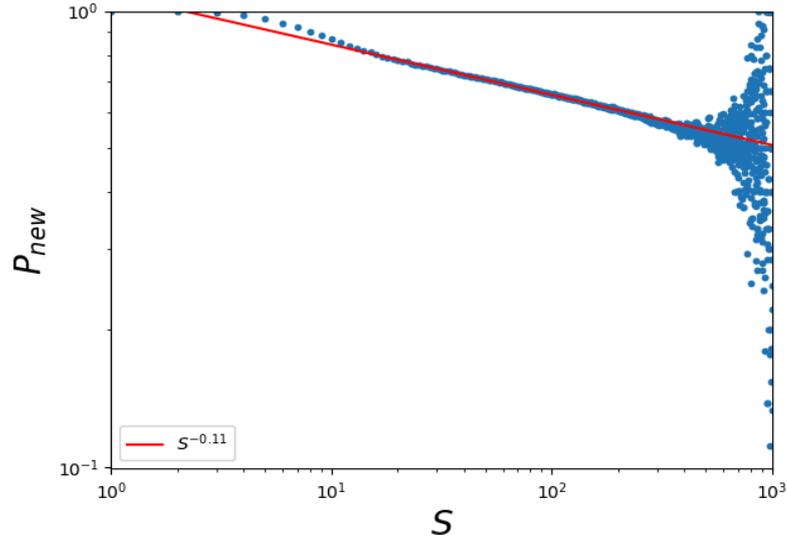

**Fig. 9** Relationship between exploration and return behavior. The horizontal axis represents the number of sites visited so far, $S$, and the vertical axis represents the probability of visiting a new site next time, $P_{new}$. Both axes are displayed on the logarithmic scale. The red line represents $S^{-0.11}$

Figure 9 illustrates the relationship between exploration and the return behavior. The horizontal axis represents the number of sites visited so far, $S$, and the vertical axis represents the probability of visiting a new site the next time, $P_{new}$. The figure shows that $P_{new} \propto S^{-0.11}$ holds over a wide range of up to several hundred steps.

## 4. Discussion

In this study, a simplified version of the agent walk model proposed by Shinohara et al. [15] was used to simulate human mobility behavior. This random-walk model generates a Cauchy walk if the base point is set to the current position of the agent. In this study, the base point was set as a fixed point, such as the home, to model human mobility behavior.

If $\gamma = 1$, the agent approaches the target point $T$ by the shortest path. Contrastingly, if $\gamma = 0$, it can be thought of as replacing a target point with a target line, as in American football when the scrimmage line is moved parallel toward the end zone of the opponent (Fig. 2). In this case, the distance between the agent and the target point may move away as a result of the move. This replacement generates conflicting movement behaviors of approaching and leaving the target point.

The model can reproduce some of the power laws found in human mobility behavior though there are no explicit rules for following scaling laws. First, as Figure 4 shows, the distance distribution can be approximated by a power distribution $P(\|\Delta X\|) \sim \|\Delta X\|^{-\alpha}$ with $\alpha \approx 1.0$. However, in the empirical data, $\alpha \approx 1.55$ [16], $\alpha \approx 1.75 \sim 2.02$ [25], and $\alpha \approx 1.80 \sim 2.16$ [22] were obtained, and longer jumps appeared more frequently in our simulation.

It can be observed from Fig. 5 that the waiting time distribution can be approximated using power distribution $P(\Delta t) \sim \Delta t^{-\beta}$ with $\beta \approx 1.44$. In empirical data, the scaling exponent concerns $\beta \approx 1.8$ in both real- and virtual-space data [16, 17], and the exponent does not match our simulation results. However, when Liu et al. [22] reanalyzed data with a higher resolution using MFR, they obtained a scaling exponent of $\beta \approx 1.57$ that is close to the simulation results in this study.

As shown in Fig. 6, Zipf's law $f_k \propto k^{-\xi}$ with $\xi \approx 1.0$ holds for the frequency of visits to each site. While $\xi \approx 1.2$ was found in empirical data for the real space [16, 23], $\xi \approx 0.94$ was obtained for the virtual space [17]. In other words, the simulation results in this study are closer to the results in the virtual space than in the real space.

Figure 7 shows that the number of agents per unit area who reside at a distance $r$ from a particular location and visit that location $f$ times in a given period can be approximated using power law $\rho(r, f) \propto (rf)^{-\eta}$ with $\eta \approx 1.9$. This scaling exponent is remarkably close to the value $\eta \approx 2.05$ obtained from empirical data [24].

For the relationship between time $t$ and number of sites visited $S(t)$, Fig. 8 shows that power law $S(t) \propto t^{\mu}$ with $\mu \approx 0.7$ holds. In the empirical data, the scaling exponent was $\mu \approx 0.6$ [16] for the real space. The scaling exponents for the virtual space were $\mu \approx 0.52$ and $\mu \approx 0.57$ for watching TV and shopping online, respectively [17]. The results are similar to our simulation results.

Thus, the model proposed in this study can reproduce by a single mechanism: both random walks, such as the Cauchy and Brownian walks found in animal migratory behavior, and the scaling laws found in human migratory behavior. The EPR model incorporates a rule of visiting a new site with probability $P_{new} = \rho S^{-\gamma}$, and the scaling exponent of $\gamma \approx 0.21$ was obtained in the empirical data [16]. As shown in Fig. 9, the $P_{new} \propto S^{-0.11}$ relationship was also established in our model as a result. It is interesting to note that although our model does not explicitly incorporate the rule to follow the scaling law, a rule similar to the EPR model generates. In the EPR model, the jump size and the waiting time are sampled from a predefined power distribution. In our model, such scaling laws are generated as a result. However, the scaling exponents do not match the empirical data. Modification of the model to fit the empirical data is a subject for future work.

# Funding

This study was supported by JSPS KAKENHI [Grant No. JP21K12009].

# Financial Interest

The authors have no relevant financial or non-financial interests to disclose.

# Author Contributions

**Shuji Shinohara:** Conceptualization, formal analysis, methodology, software, writing, original draft preparation, and funding acquisition. **Daiki Morita:** Software, review, and editing. **Nobuhito Manome:** Writing, reviewing, and editing. **Hayato Hirai:** Software, review, and editing. **Ryosuke Kuribayashi:** Software, review, and editing. **Toru Moriyama:** Writing, reviewing, and editing the manuscript. **Yoshihiro Nakajima:** Writing, reviewing, and editing. **Pegio-Yukio Gunji:** Writing, reviewing, editing, and supervision. **Ung-il Chung:** Writing, review, editing, supervision, and project administration.

## Data Availability

The datasets generated during and/or analysed during the current study are available in the GitHub repository, https://github.com/shinoharaken/Human-Mobility.